\documentclass[aps,prl,reprint,superscriptaddress, floatfix]{revtex4-1}

\usepackage{siunitx}
\usepackage{physics}
\usepackage[capitalize]{cleveref}
\usepackage{graphicx}
\makeatletter
\let\saved@includegraphics\includegraphics
\AtBeginDocument{\let\includegraphics\saved@includegraphics}
\renewenvironment*{figure}{\@float{figure}}{\end@float}
\makeatother
\usepackage{color}
\usepackage{xr}

\usepackage{amsmath}
\usepackage{soul}

\begin{document}

\title{Non-Classical Longitudinal Magneto-Resistance in Anisotropic Black Phosphorus} 

\author{F. Telesio}
\affiliation{NEST, Istituto Nanoscienze-CNR and Scuola Normale Superiore, I-56127 Pisa, Italy}
\author{N. Hemsworth}
\affiliation{Department of Electrical and Computer Engineering, McGill University, Montr\'eal, Qu\'ebec, H3A 2A7, Canada}
\author{W. Dickerson}
\affiliation{Department of Electrical and Computer Engineering, McGill University, Montr\'eal, Qu\'ebec, H3A 2A7, Canada}
\author{M. Petrescu}
\affiliation{Department of Physics, McGill University, Montr\'eal, Qu\'ebec, H3A 2T8, Canada}
\author{V. Tayari}
\affiliation{Department of Electrical and Computer Engineering, McGill University, Montr\'eal, Qu\'ebec, H3A 2A7, Canada}
\author{Oulin Yu}
\affiliation{Department of Physics, McGill University, Montr\'eal, Qu\'ebec, H3A 2T8, Canada}
\author{D. Graf}
\affiliation{National High Magnetic Field Laboratory, Tallahassee, FL 32310, United States}
\author{M. Serrano-Ruiz}
\affiliation{Istituto Chimica dei Composti OrganoMetallici-CNR, I-50019 Sesto Fiorentino, Italy}
\author{M. Caporali}
\affiliation{Istituto Chimica dei Composti OrganoMetallici-CNR, I-50019 Sesto Fiorentino, Italy}
\author{M. Peruzzini}
\affiliation{Istituto Chimica dei Composti OrganoMetallici-CNR, I-50019 Sesto Fiorentino, Italy}
\author{M. Carrega}
\affiliation{NEST, Istituto Nanoscienze-CNR and Scuola Normale Superiore, I-56127 Pisa, Italy}
\author{T. Szkopek}
\affiliation{Department of Electrical and Computer Engineering, McGill University, Montr\'eal, Qu\'ebec, H3A 2A7, Canada}
\author{S. Heun}
\affiliation{NEST, Istituto Nanoscienze-CNR and Scuola Normale Superiore, I-56127 Pisa, Italy}
\author{G. Gervais}
\affiliation{Department of Physics, McGill University, Montr\'eal, Qu\'ebec, H3A 2T8, Canada}

\date{\today}

\begin{abstract}
Resistivity measurements of a few-layer black phosphorus (bP) crystal in parallel magnetic fields up to 45~T are reported as a function of the angle between the in-plane field and the source-drain (S-D) axis of the device.  The crystallographic directions of the bP crystal were determined by Raman spectroscopy, with the zigzag axis found within $5^{\circ}$ of the S-D axis, and the armchair axis in the orthogonal planar direction.  A transverse magneto-resistance (TMR) as well as a classically-forbidden longitudinal magneto-resistance (LMR) are observed. Both are found to be strongly anisotropic and non-monotonic with increasing in-plane field.  Surprisingly,  the relative magnitude (in \%) of the  {\it positive} LMR is larger than the TMR  above $\sim 32$~T.  Considering the known anisotropy of bP whose zigzag and armchair effective masses differ by a factor of approximately seven, our experiment strongly suggests this LMR to be a consequence of the anisotropic Fermi surface of bP.

\end{abstract}

\maketitle


\section{Introduction}
Magnetoresistance (MR) is a phe\-nom\-e\-non in which a ma\-te\-rial's resistivity increases or decreases due to the presence of a magnetic field $\bf B$. Transport measurements typically require the presence of an electric field ${\bf E}$ so as to establish an average current with charge velocity ${\bf v}$ along a preferred direction. The  overall force on the charge carriers $q$  (electrons or holes) is  simply given by the Lorentz force,   ${\bf F}=q({\bf E} +  {\bf v} \times {\bf B})$. From this stems two important limiting cases: one where the  current flow is perpendicular to {\bf B} and for which the magnitude contribution of the Lorentz force is maximal, and the other where it is parallel to {\bf B} and there is no magnetic contribution to the Lorentz force.  This simple classical picture therefore implies that MR is forbidden in the latter case, however decades of research have shown that a material can develop a longitudinal magnetoresistance (LMR) when the current and the magnetic field are parallel. The exact set of conditions for which a non-classical LMR {\it can or cannot} be observed remains a highly debated topic which has gained renewed interest recently within the context of Weyl semimetals and topological insulators \cite{Son2013,Goswami2015,Andreev2018}.

Pal and Maslov \cite{Pal2010} have studied theoretically the non-classical LMR on generic grounds and they proposed a set of necessary and sufficient conditions within the context of Fermi surface (FS) morphology for a {\it three-dimensional} system. While not all anisotropy leads to a LMR, it was shown that an {\it angular} anisotropy of the FS along the magnetic field direction is a {\it sufficient} condition. Few-layer black phosphorus (bP) provides a key system for this since its Fermi surface (FS) is highly anisotropic with effective masses for holes $m_{ac} = 0.11 m_0$ and $m_{zz} = 0.71 m_0$ ($m_0$ is the bare electron mass) along the armchair $(ac)$ and zigzag $(zz)$ directions, respectively \cite{Qiao2014}. LMR  has been observed previously in bulk crystals of bP  \cite{Strutz1994,Hou2016} yielding only limited progress in its understanding.  This is the subject of this work, where an experiment was designed to perform magneto-transport measurements in a few-layer bP device in the presence of a {\it purely parallel magnetic field} that could be rotated {\it in the plane of the bP few-layer flake}, and up to 45~T field.  A strong classically-forbidden LMR was found whose non-monotonic field dependence closely matches a parabolic behavior. Even though LMR has been studied for decades in three-dimensional systems, to our knowledge there is no rigorous theory in two-dimensional anisotropic systems and as such our results are calling for future theoretical work in this direction.

\section{Results}

\subsection{Device structure} 

The geometry of the few-layer field-effect transistor device is shown in figure~\ref{fig:1}(a). Details on device fabrication are provided in the supplementary information (SI). The flake thickness was carefully determined from the atomic force microscopy (AFM) measurements shown in the SI, and is $t = (16 \pm 1)$~nm in the channel region. The channel length and average weighted width are $L = 25.4$~$\mu$m and $W = 4.4$~$\mu$m, respectively. The transistor includes a conventional back gate and two additional top gates labeled TG1 and TG2 in figure~\ref{fig:1}(a) employing a combination of PO$_x$ and Al$_2$O$_3$ as oxide dielectrics \cite{Dickerson2018}. For the measurements presented here, TG1 is not used. We define $\varphi$ as the angle between the source-drain (S-D) axis of the device and the in-plane magnetic field used in the experiment (see figure~\ref{fig:1}(a) and figure~\ref{fig:2}(a)). With this convention, at  $\varphi=0^{\circ}$ the magnetic field is parallel to the S-D axis and hence to the current (${\bf B}\parallel {\bf I}$), whereas at $\varphi = \pm 90^{\circ}$ the magnetic field is perpendicular to  the S-D axis and the current (${\bf B}\perp {\bf I}$). For clarity, the same convention has been used in displaying the polarized Raman measurements shown in figure~\ref{fig:1}(c) and (d).

\begin{figure}[tb]
   \includegraphics[width=0.95\columnwidth]{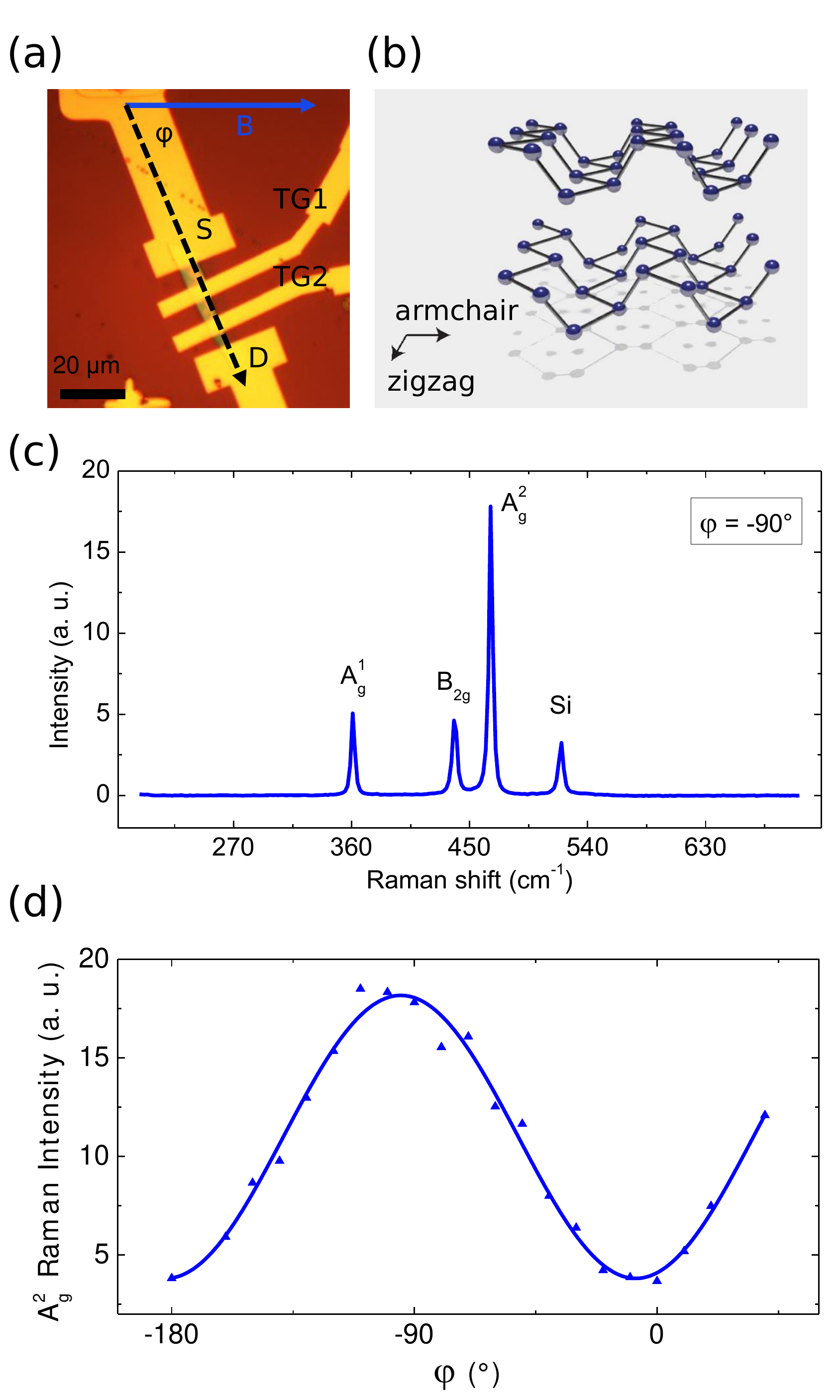}
   \caption{\label{fig:1}({\bf a}) Optical microscopy image of the device with labeling of source (S) and drain (D) contacts, top gates TG1 and TG2, as well as the definition of the angle $\varphi$ as the angle between source-drain and the magnetic field axis.  ({\bf b}) The puckered crystal structure of black phosphorus with armchair and zigzag directions indicated. ({\bf c}) Raman spectrum of the bP device, with baseline subtracted. The spectrum was acquired at $\varphi = -90^\circ$, {\it i.e.} with light polarization perpendicular to the device axis. The three bP Raman peaks are labeled, as well as the peak emanating from the  Si substrate. ({\bf d}) Polarized Raman measurement. $A^2_g$ intensity as a function of incoming laser light polarization angle $\varphi$. The maximum of $A^2_g$ is along the armchair axis, the minimum at $(-5 \pm 2)^{\circ}$ along zigzag, and therefore the S-D axis of the device is approximately aligned with the zigzag direction of the flake (see SI for additional details).}
\end{figure}

\subsection{Raman characterization} 

The crystal orientation with respect to the source-drain  contacts axis was determined via polarized Raman spectroscopy. Here, the crystallographic orientation of the bP crystal, shown in figure~\ref{fig:1}(b), was determined by measuring the Raman peak intensities associated with the in-plane vibrational modes of black phosphorus (A$^2_g$ and B$_{2g}$) \cite{Sugai1985} with respect to the linear polarization of the incident laser \cite{Chen2017,Wang2015a}. In particular, the maximum of the A$^2_g$ mode corresponds to the armchair direction \cite{Sugai1985,Kim2015a,Phaneuf-LHeureux2016a}. These Raman data are shown in figure~\ref{fig:1}(d) as a function of polarization angle $\varphi$, whereby  for $0^\circ$ the polarization of the incoming laser beam is parallel to the axis of the bP FET channel (dashed line in figure~\ref{fig:1}(a)). From these data, we determine the bP crystal orientation to be such that the zigzag direction is at angle $\theta = (-5 \pm 3)^{\circ}$ from the S-D channel axis, as can be seen in figure~\ref{fig:1}(d) (see also SI).

\subsection{Magneto-transport measurements} 

\begin{figure}[tb]
   \includegraphics[width=\columnwidth]{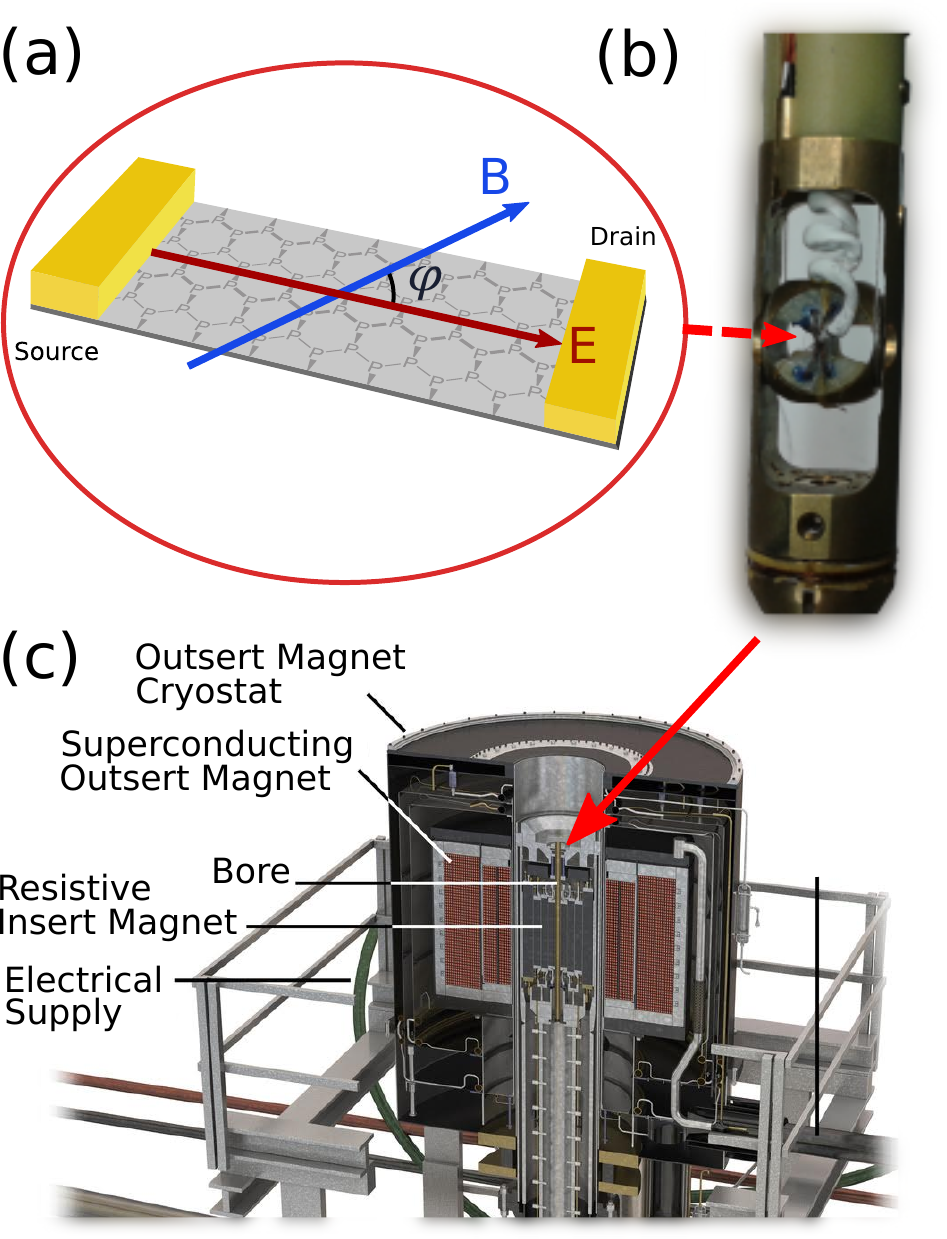}
   \caption{\label{fig:2}({\bf a}) Schematics of the device with electric field ${\bf E}$ and magnetic field ${\bf B}$ indicated, as well as crystallographic orientation. The angle $\varphi$ is defined as the angle between source-drain and magnetic field axis. ({\bf b}) Photo of the rotator probe as mounted in the cryostat insert. The sample is rotating with the magnetic field in the bP plane. ({\bf c}) Schematics of the hybrid magnet of National High Magnetic Field Laboratory used for the high-field measurements. The outer, superconducting  magnet provides a field in the 0~T to 11.4~T range, and together with the resistive insert a field of 45~T can be reached.}
\end{figure}

\begin{figure}[tb]
   \includegraphics[width=\columnwidth]{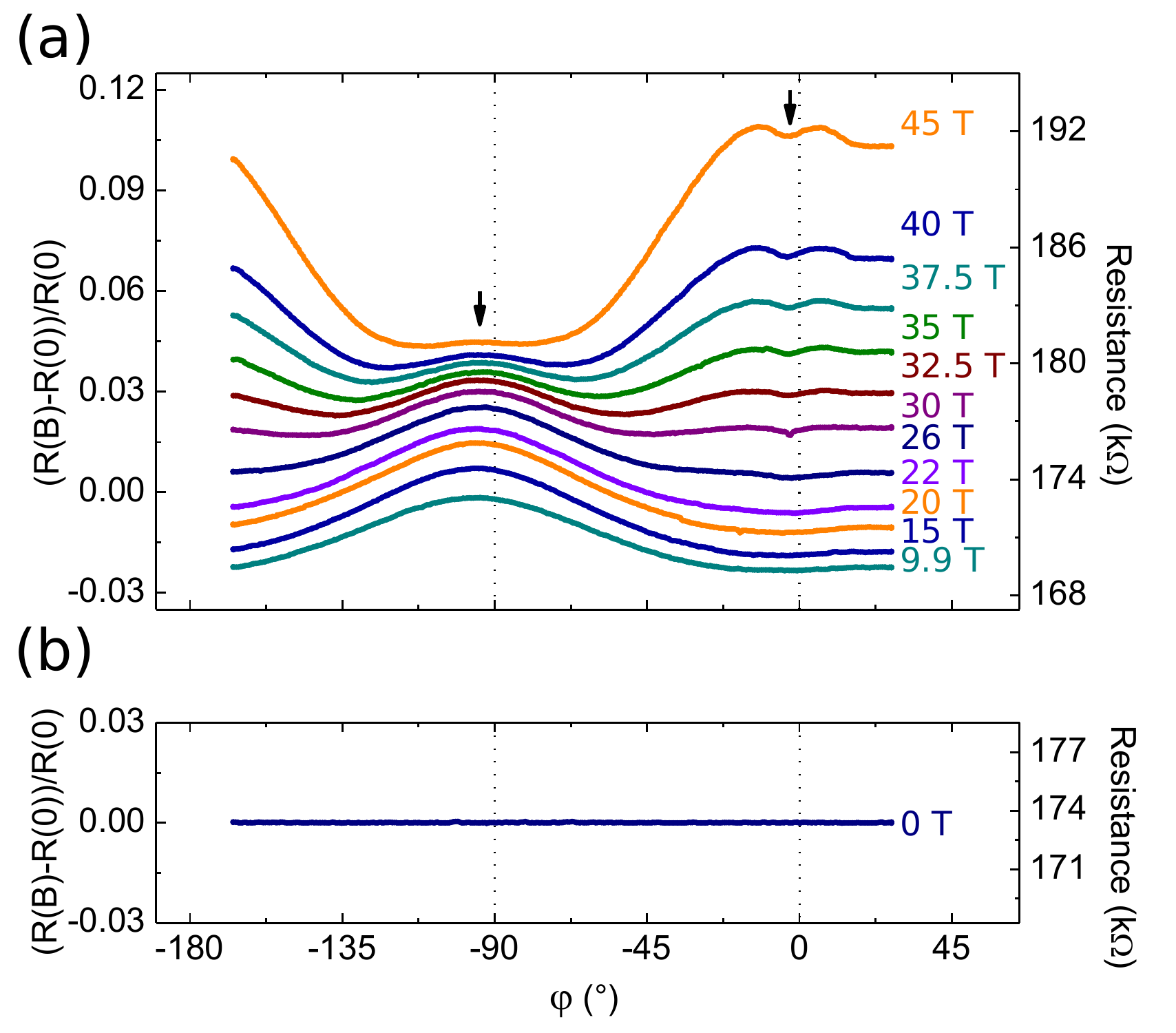}
   \caption{\label{fig:3}({\bf a}) Magneto-resistance, defined as $(R(B) - R(0)) / R(0)$, {\it vs}. the in-plane angle of rotation  $\varphi$ at various magnetic fields,  left axis. The resistance value is displayed on the right axis. ({\bf b}) Zero-field resistance value measured versus $\varphi$. In both panels, the data was taken at 323 mK temperature.}
\end{figure}

During  low--tem\-per\-a\-ture magneto-transport measurements, the gate volt\-age dependence revealed an inherent p-type character for the bP FET. The device exhibited an intrinsic carrier concentration $n = 2.2 \times 10^{12}$~cm$^{-2}$  and a field-effect mobility $\mu = 83$~cm$^2$/(Vs) at $T = 1.64$~K and $B=11.4$~T, as provided in the SI. In all displayed measurements, except when explicitly stated, TG1 and the back gate were kept to ground, whereas TG2 was set to $-1$~V to increase the bP conductance. The device was mounted on a calibrated step-by-step rotator (shown in figure~\ref{fig:2}(b)) that could rotate the sample with the magnetic field axis in the plane of the device. The rotator had an angular resolution  within  $0.02^{\circ}$ which is much less than the systematic errors on $\varphi$ arising from the crystallographic orientation measurements of bP by Raman spectroscopy. The experiment was performed at low temperatures down to 300~mK and up to a 45~T magnetic field in the hybrid magnet of the National High Magnetic Field Laboratory in Tallahassee, shown in figure~\ref{fig:2}(c). The in-plane magneto-transport was investigated by performing several angular sweeps at various magnetic fields,  as well as magnetic field sweeps at various angles, to check for data consistency. The normalized magneto-resistance of the device, defined as $\Delta R/R=(R(B)-R(0))/R(0)$, with $R(0) \equiv R(B=0)= 173.3$~k$\Omega$, is  shown for various magnetic fields in figure~\ref{fig:3} on the left axis of the graph, whereas the right axis displays the resistance values. The zero-field resistance $R(0)$ is also shown in the lower panel of the same figure. Overall, the MR is observed to have a strong dependence on the angle $\varphi$ and varies non-monotonically with the increasing in-plane magnetic field.

Upon closer inspection of figure~\ref{fig:3} it becomes apparent that the maxima and minima in MR are not {\it exactly} aligned at $\varphi=0^\circ$ and $90^\circ$ (as indicated by the dotted vertical lines in the figure) but instead are slightly shifted to a lower value (as indicated by the arrows). Tracking the position of the maximum in MR close to $-90^{\circ}$ and the minimum close to $0^{\circ}$, their positions are found to be nearly constant at all magnetic fields, and given by  $(-94.4 \pm 2.0)^{\circ}$ and $(-2.8 \pm 2.0)^{\circ}$, respectively. The $\pm~ 2^{\circ}$ quoted here was estimated from the possible orientation error in mounting the device on the high magnetic field sample holder.  This significant deviation from the precise values of the device axis suggest that neither the device axis nor the current direction define the relevant coordinate system for the MR. In fact, these angles correspond within error to the orientation of the bP crystal and the directions of zigzag and armchair axes of the bP crystal, indicating that these are the physically-relevant directions determining the magneto-resistance.

Figure~\ref{fig:4} displays cross-sections of figure~\ref{fig:3} for the parallel ($\varphi=0^{\circ}$) and perpendicular ($\varphi=-90^{\circ}$) configurations of the magnetic field with respect to the current direction. The longitudinal magneto-resistance (${\bf B}\parallel {\bf I}$) is first negative, passes through a minimum at approximately 11~T, and then increases to reach a positive value at approximately 26~T. The LMR is found to take  larger value than the transverse magneto-resistance (${\bf B}\perp {\bf I}$) at a crossover field of approximately 32.5~T. Remarkably, the magnetic field dependence of the LMR can be well described by a shifted parabola centred at $B=(11.9\pm 0.3)$~T (red line).  In contrast, the TMR is nearly constant up to 11~T field (where the LMR reached its minimum value), and then increases roughly linearly with the in-plane magnetic field increasing.

Resistance measurements taken at fixed angle as a function of magnetic field and shown in figure~S4(d) of the SI  confirm this trend. At angles $\varphi$ close to $0^{\circ}$ and $-90^{\circ}$, the data display a similar crossover, whereas at $\varphi \approx -180^{\circ}$ and  $0^{\circ}$ the data highlight the $180^{\circ}$ periodicity of the phenomenon. Finally, upon reversal of the magnetic field, the magneto-resistance showed no significant dependence on the field direction, see SI figure~S4(e), except for a slight overall shift that is attributed to a small change in carrier density occurring during a temperature cycling of the device from base to $\sim 40$~K temperatures. Since the transport measurements were performed in a two-point configuration, the robustness of the observed LMR and TMR against magnetic field  polarity ensures that solely the in-plane resistivity played a role in the experiment.

\section{Discussion}

\subsection{Transport regime and length scales} To establish the regime of charge transport in the bP flake, we compare the relevant length scales obtained from the experimentally determined carrier density $n$ and mobility $\mu$. We apply a two--dimensional Drude model justified here by a self-consistent Schr\"odinger-Poisson calculation showing that the charge density in bP flakes on SiO$_2$ is concentrated within a surface accumulation layer with mean thickness $\langle z \rangle \simeq 3$~nm \cite{Tayari2015}. Within the effective mass approximation, the hole dispersion is given by $E = \hbar^2 k_{zz}^2 / 2m_{zz} + \hbar^2 k_{ac}^2 / 2m_{ac}$, quantifying the anisotropy in charge carrier motion along zigzag and armchair directions.

The Fermi wavevector is itself anisotropic, with the relevant Fermi wavevector being $k_{F,zz}$ for charge carriers moving in the zigzag direction with current flow in the bP flake. The wavevector is $k_{F,zz}= (m_{zz}/m_{ac})^{1/4} (2\pi n)^{1/2} = 0.59~\mathrm{nm}^{-1}$, which corresponds to a Fermi wavelength $\lambda_{F,zz} = 2 \pi / k_{F,zz}= 10.6~\mathrm{nm}$. The Fermi wavelength is larger than the mean thickness $\langle z \rangle$ of the accumulation layer, and smaller than the flake thickness $t$. The Fermi velocity $v_{F,zz}$ along the zigzag axis is $v_{F,zz} = \hbar k_{F,zz} / m_{zz} = 9.7 \times 10^4 ~\mathrm{ms}^{-1}$. The elastic
scattering time $\tau_{zz} = m_{zz} \mu/e = 3.3\times10^{-14}~\mathrm{s}$ and the elastic mean free path $\ell_{e,zz}=v_{F,zz} \tau_{zz}= 3.2$~nm \cite{Ando1982}, where the mobility $\mu=\mu_{zz}$ was measured for current flow along the zigzag axis. The mean free path $\ell_{e,zz}$ is similar to the thickness of the accumulation layer $\langle z \rangle$.

The Ioffe-Regel criterion for localization \cite{Gurvitch1981,Lee1985} can be used to ascertain the regime of charge transport. We find $k_{F,zz} \ell_{e,zz} = 1.9$, indicating that transport within our bP device is diffusive, albeit close to the crossover from diffusive to localized transport ($k_{F} \ell_{e} = 1$). We can compare the length scales of charge carrier transport with the magnetic length $\ell_B = \sqrt{\hbar / eB}$, which decreases from 11.4 nm at $B =$~5 T to 3.8 nm at $B =$~45 T. Only at the highest fields used in our experiments does the magnetic length approach the thickness of the accumulation layer.

\begin{figure}[tb]
   \includegraphics[width=\columnwidth]{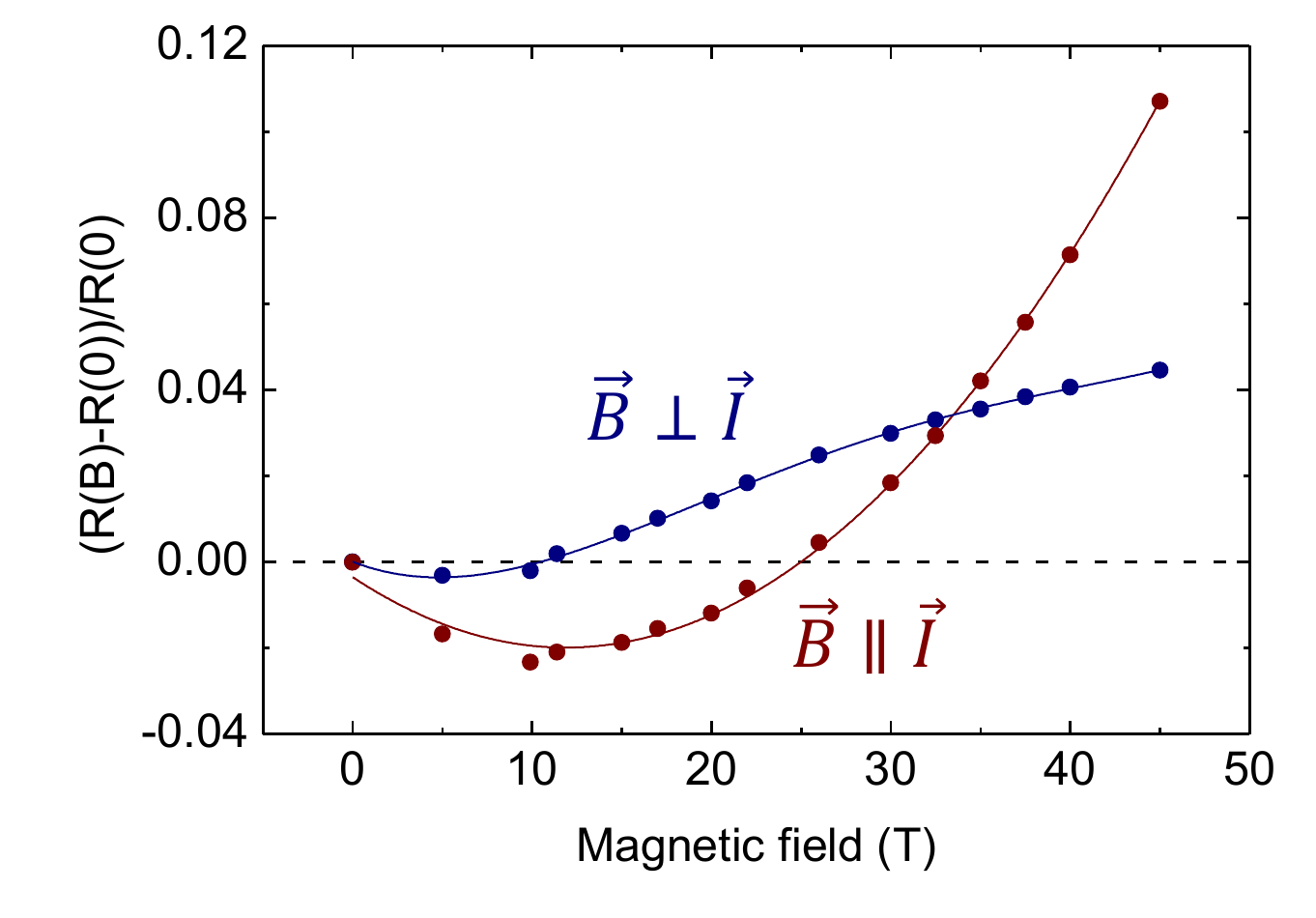}
   \caption{\label{fig:4} Magnetic-field dependence of the LMR and TMR  showing distinct behaviours. It is based on the data shown in figure~3(a). The red line is a fit of the LMR (${\bf B}\parallel {\bf I}$) data with a shifted parabola centred at $B=(11.9\pm 0.3)$~T. The blue line is a guide-to-eye for the TMR data (${\bf B}\perp {\bf I}$).}
\end{figure}

\subsection{Spin-orbit scattering} 

A negative magneto-resistance in the strongly localized regime has been observed before, and has been ascribed to the orbital MR due to quantum-interference among random paths in the hopping process \cite{Sivan1988,Faran1988}. In particular, it was reported that an anisotropy in the MR is indicative of this magneto-orbital mechanism and that this orbital MR is \textit{always} negative \cite{Faran1988}.  Later work suggested that inclusion of spin-orbit scattering in the theory would lead to a positive MR \cite{Pichard1990,Hernandez1992}. Such positive MR in the strongly localized regime has been observed in many disordered materials \cite{Jiang1994}, however this reasoning is unlikely to be applicable here given that spin-orbit coupling in bP is so weak \cite{Qiao2014,Popovic2015}. Besides, the in-plane anisotropy of the $g$-factor of bP is also negligible $(g_{ac} \approx g_{zz})$ \cite{Li2016,Zhou2017} and close to the bare electron value $g = 2$, indicative of negligible exchange enhancement \cite{Li2016}.

\subsection{Anisotropic Fermi surface} 

Non-classical LMR has been observed in a variety of systems ranging from three--dimensional metals, topological materials, and semiconductors (see \cite{Goswami2015} for a recent discussion). Already in the 1960's this phenomenon was attracting interest in the context of metals with non-trivial Fermi surfaces \cite{Pippard1964}, and subsequently in semiconductors with non-parabolic bands \cite{Phadke1975}. More recently, it has been discussed for massless Dirac and Weyl fermionic systems \cite{Son2013,Goswami2015,Sekine2017}. In 2010,  Pal and Maslov \cite{Pal2010} derived  the necessary and {\it sufficient} conditions for the occurrence of LMR within the context of Fermi surface morphology of a three--dimensional electronic system;  while they have shown that an anisotropic Fermi surface is a prerequisite, not all types of anisotropy will give rise to the effect. They specified that an angular anisotropy of the Fermi surface along the magnetic field direction is a  {\it sufficient} condition in three dimensions. Our case,  for which the electronic system is defined by a two-dimensional hole gas confined at the bP/SiO$_2$ interface, provides an interesting example of anisotropic FS since the bP  Fermi surface is elliptical and its effective masses in the armchair and zigzag directions differ by a factor of approximately seven. 

While Pal and Maslov studied the conditions for LMR to occur on generic grounds for a three--dimensional system, a quantitative theoretical description for LMR in two--dimensional systems with anisotropic FS is still lacking. Moreover, Goswami \textit{et al.}~\cite{Goswami2015}~state that for a 3D metal the existence of a finite LMR is purely a quantum-mechanical effect and a direct consequence of the axial anomaly. Furthermore, they showed that in the presence of both neutral and ionic impurities, the LMR becomes first negative for low fields, and then positive for high fields  after passing through a minimum, which is reminiscent of what is observed in the LMR data shown in figure~\ref{fig:4}. We caution, however, that their theory was developed for the quantum limit  where $\omega_c \tau \gg 1$, with $\omega_c$ the cyclotron frequency and $\tau$ the elastic scattering time. Nevertheless, the authors of \cite{Goswami2015} underline the good agreement of their results with calculations of Spivak and co-workers in the semi-classical \cite{Son2013} and diffusive regime \cite{Andreev2018}, both valid for $\omega_c \tau \ll 1$.  These recent theoretical advances suggest that a similar framework should be developed in 2D systems, especially given the present interest in 2D atomic crystals of all kinds. For instance, it is known that applying a purely parallel magnetic field on a 2DEG can lead to a magneto-orbital coupling sufficient to generate a substantial magneto-resistance, as proposed theoretically by Das Sarma and Hwang \cite{DasSarmaS2000} and found experimentally in GaAs 2DEGs with finite widths \cite{zhou2010}. Black phosphorus being a simple system to test the effects of FS anisotropy, it is in our view likely that FS anisotropy could lead to the MR modulation found in this work, even though no quantitative theory in 2D exists at the moment.

\section{Conclusions} The classically-forbidden LMR has been studied in different materials systems for decades and only recently the conditions under which it  can be observed were explored theoretically in three dimensions. One clear route towards an LMR  in 3D involves an anisotropic Fermi surface, and in its simplest expression an anisotropic FS in 2D can be thought out of an ellipse.  Our work on black phosphorus at high-magnetic field provides an important example for the appearance of LMR in a two-dimensional anisotropic system, and such observation of a longitudinal magneto-resistance for all fields inspected is confirmed. This LMR was discovered to be strongly anisotropic, non-monotonic,  positive beyond $\sim 26$~T  and dominant in strength beyond $\sim 30$~T field. Even though the magnetic length at the highest magnetic field used here (45~T) is more than one order of magnitude larger than the phosphorus bond length, this LMR work demonstrates once more the potential generated by high magnetic fields in the understanding of band structure effects on charge transport in atomic crystals. Our experimental data strongly suggests that anisotropy most likely plays an important role in the appearance of LMR, thereby calling for theoretical descriptions of magneto-transport properties in the presence of anisotropic Fermi surfaces to be developed.

\section*{Acknowledgements}
We are grateful to Danil Bukhvalov, Alina Mre\'nca--Kolasi\'nska, and Bart\l{}omiej Szafran for helpful discussions. This work was funded by NSERC (Canada), CIFAR, FRQNT (Qu\'ebec), and the CRC program (Thomas Szkopek and Guillaume Gervais). A portion of this work was performed at the National High Magnetic Field Laboratory which is supported by NSF Cooperative Agreement No. DMR-1157490, and the State of Florida. Francesca Telesio,
Manuel Serrano-Ruiz, Maria Caporali, Maurizio Perruzini, and Stefan Heun thank the European Research Council for funding the project PHOSFUN {\it Phosphorene functionalization: a new platform for advanced multifunctional materials} (Grant Agreement No. 670173) through an ERC Advanced Grant to Mauricio Peruzzini. Francesca Telesio   acknowledges financial support by CNR-Nano through the SEED project SURPHOS. Stefan
Heun acknowledges support from Scuola Normale Superiore, Project No. SNS16 B HEUN-004155. Matteo Carrega acknowledges support from the bilateral project CNR/Royal Society number IES/R3/170252.

%
\newpage
\begin{widetext}
\appendix{\textit{Supporting Information for} Non-Classical Longitudinal Magneto-Resistance in Anisotropic Black Phosphorus\\ }
\renewcommand{\thefigure}{S\arabic{figure}}
\setcounter{figure}{0}

\section{Black Phosphorus Synthesis}

Black phosphorus (bP) crystals were synthesized from red phosphorus using a published procedure~\cite{Kopf2014}, wherein high purity red phosphorus ($>99.99\%$), tin ($>99.999\%$), and gold ($>99.99\%$) are heated in a muffle oven with $\text{SnI}_4$ catalyst. The solid product was placed in a quartz tube, subjected to several evacuation-purge cycles with $\text{N}_2$ gas, and then sealed under vacuum. The evacuated quartz tube was heated to $406\,\si{\degreeCelsius}$ at $4.2 \,\si{\degreeCelsius}/\si{\min}$, where it remained for 2 hours. The tube was then heated to $650\,\si{\degreeCelsius}$ at $2.2\,\si{\degreeCelsius}/\si{\min}$ and held at this temperature for 3 days. The tube was then cooled to room temperature at $0.1\,\si{\degreeCelsius}/\si{\min}$. The final product is crystalline bP with a typical size of a few mm.

\section{Device fabrication}

Black phosphorus FETs were prepared by mechanical exfoliation of bulk bP crystals onto a degenerately doped Si wafer with a $300\,\si{\nano\meter}$ $\text{SiO}_2$ layer prepared by dry thermal oxidation. Exfoliation was performed in a nitrogen glove box to suppress photo-oxidation, and the $\text{SiO}_2$ surface was treated with a hexamethyldisilazane (HMDS) layer to suppress charge transfer doping. Standard electron beam lithography (EBL) was used to define $5\,\si{\nano\meter}$ Ti/$80\,\si{\nano\meter}$ Au metal electrodes. A phosphorus oxide layer was then formed by $\text{O}_2$ plasma etching the sample for 3 minutes with a $300\,\si{\watt}$ RF bias power and 10 standard cubic centimeters per minute $\text{O}_2$ flow at $200\,\si{\milli}\text{Torr}$. A $3\,\si{\nano\meter}$ $\text{Al}_2\text{O}_3$ layer was deposited atop the bP by atomic-layer deposition (ALD) at $150\,\si{\degreeCelsius}$ through an EBL-defined window, following a previously reported procedure \cite{Dickerson2018}. A top-gate metal layer was defined by a further EBL step followed by metal deposition ($5\,\si{\nano\meter}$ Ti/$80\,\si{\nano\meter}$ Au).

\section{Determination of the Flake Thickness}

An atomic force microscope (AFM) was used to de\-ter\-mine the thickness of the bP flake. The results are shown in figure~\ref{Sfig:AFM}. These measurements were performed after the transport measurements were completed to avoid potentially damaging the surface. AFM was performed in a nitrogen glovebox with a NanoSurf FlexAFM, in tapping mode, using an N-type silicon tip with a radius of $<10$~$\si{\nano\meter}$.

\begin{figure}[tb]
   \includegraphics[scale=0.7]{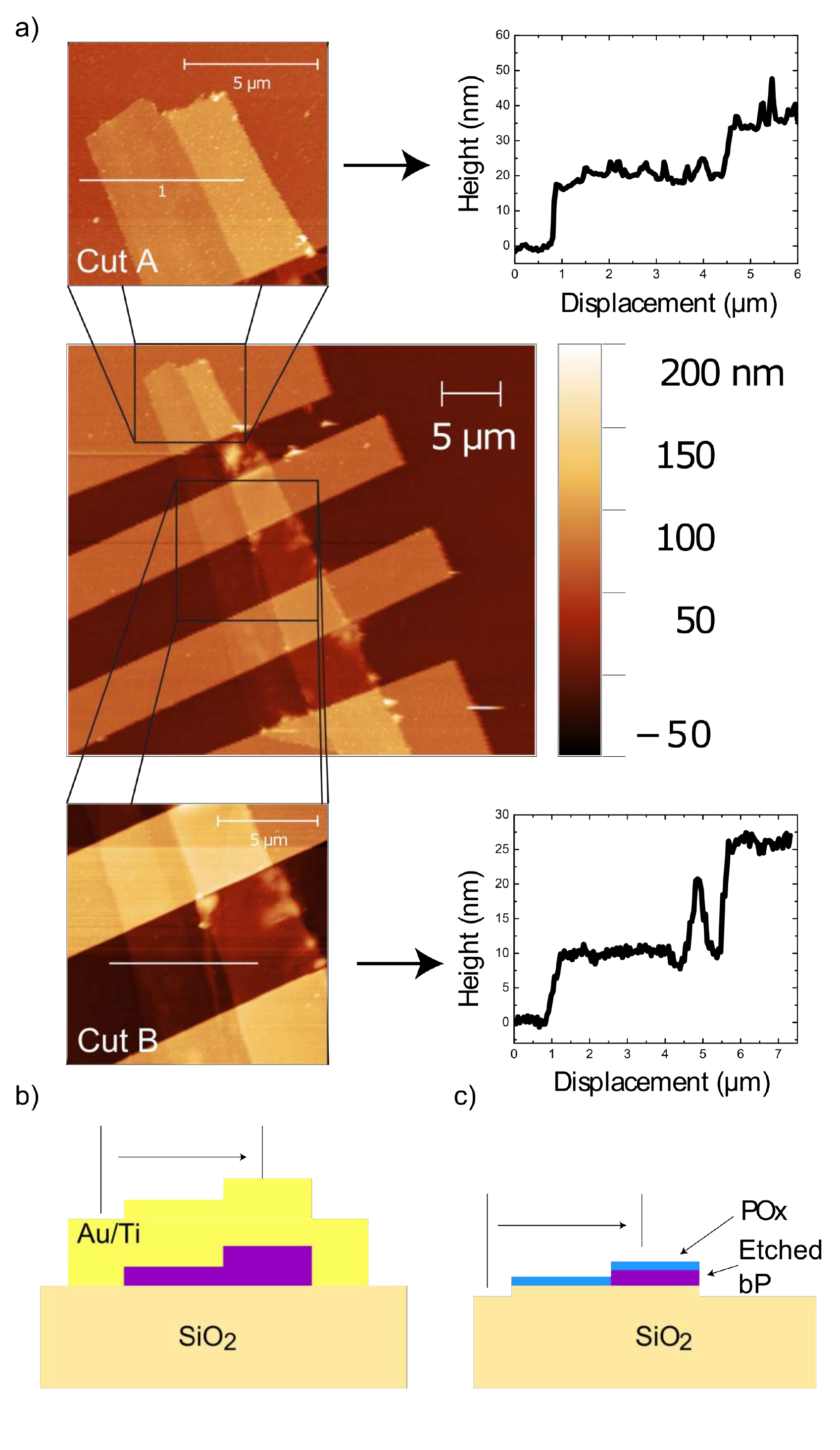}
   \caption{\label{Sfig:AFM}({\bf a}) Atomic Force Microscopy (AFM) image of the device. Two zoomed-in images are included together with their cross--sectional profiles (labelled cut A and cut B). The cross-sections were taken along the white lines. The upper panel, cut A, contains a trace over one of the contacts. As the contact was deposited before the etching process, the height of the protruding gold is equal to the thickness of the buried bP flake before etching. The lower panel, cut B, depicts etched bP. Cross sectional schematics are depicted in ({\bf b}) and ({\bf c}) to further clarify the AFM measurements with ({\bf b}) and ({\bf c}) corresponding to cut A and cut B, respectively. The vertical lines indicate the beginning and the end of the AFM cross sections.}
\end{figure}

Because of the multi-step fabrication process, a careful evaluation of flake thickness is necessary. Source and drain contacts were evaporated on the flake before the plasma etching process, which consumes phosphorus and produces a PO$_x$ layer on top of the flake. Evaluating the thickness from the step of the gold layer in the source or drain electrode region therefore provides information on the original thickness of the flake before the etching process, as depicted in figure~\ref{Sfig:AFM}~(b). This region is shown in the upper enlarged image of figure~\ref{Sfig:AFM}~(a), cut A. The height estimation along cut A shows a double tiered flake with a thinner terrace of $(20.24 \pm 0.42$)~$\si{\nano\meter}$ and a thicker terrace of $(36.14 \pm 0.93)$~$\si{\nano\meter}$, giving a height difference between the two of $(15.90 \pm 1.02)$~$\si{\nano\meter}$.

Charge transport between source and drain contacts occurs, on the other hand, in the etched region, the thickness of which can be determined by taking an AFM cross-section in the region between the two top gates (lower zoom-in and cut B in figure~\ref{Sfig:AFM} (a)). In this region as well, two distinct tiers were observed, related to the thinner ($(10.09\pm 0.16)$~$\si{\nano\meter}$) and thicker ($(26.04\pm0.30)$~$\si{\nano\meter}$) parts of the flake. Again, the height difference is $(15.95 \pm 0.34)$~$\si{\nano\meter}$. From the calibration of the reactive ion etching process \cite{Dickerson2018} we know that the thin part of the flake is fully composed of PO$_x$, as shown in the sketch in figure~\ref{Sfig:AFM}~(c). Therefore, the bP flake thickness equals the height difference between the flake tiles, which is approximately $(16 \pm 1)$~$\si{\nano\meter}$.

\section{Polarized Raman Characterization}

Polarized Raman spectroscopy was performed using a Renishaw inVia system equipped with a 532 nm laser, a half-wavelength retarder (half-wave plate), and a motorized stage for 2D mapping of samples. A laser spot size of approximately $1\,\si{\micro\meter}$ in diameter was used. Polarized Raman spectroscopy was performed keeping the sample in a fixed position and rotating  through the half-wavelength retarder the orientation of the polarization of the incoming light. The detector, placed in backscattering configuration, was not polarization selective and acquired the full signal coming from the sample. A diagram of the Raman setup is shown in figure~\ref{Sfig:Raman-a}. In order to monitor degradation and possible mechanical shifts of the sample during measurement, optical images of the device were acquired before each spectrum. No significant change in morphology and position of the sample between the beginning and the end of the measurements was observed.

\begin{figure}[tb]
   \includegraphics[scale=0.7]{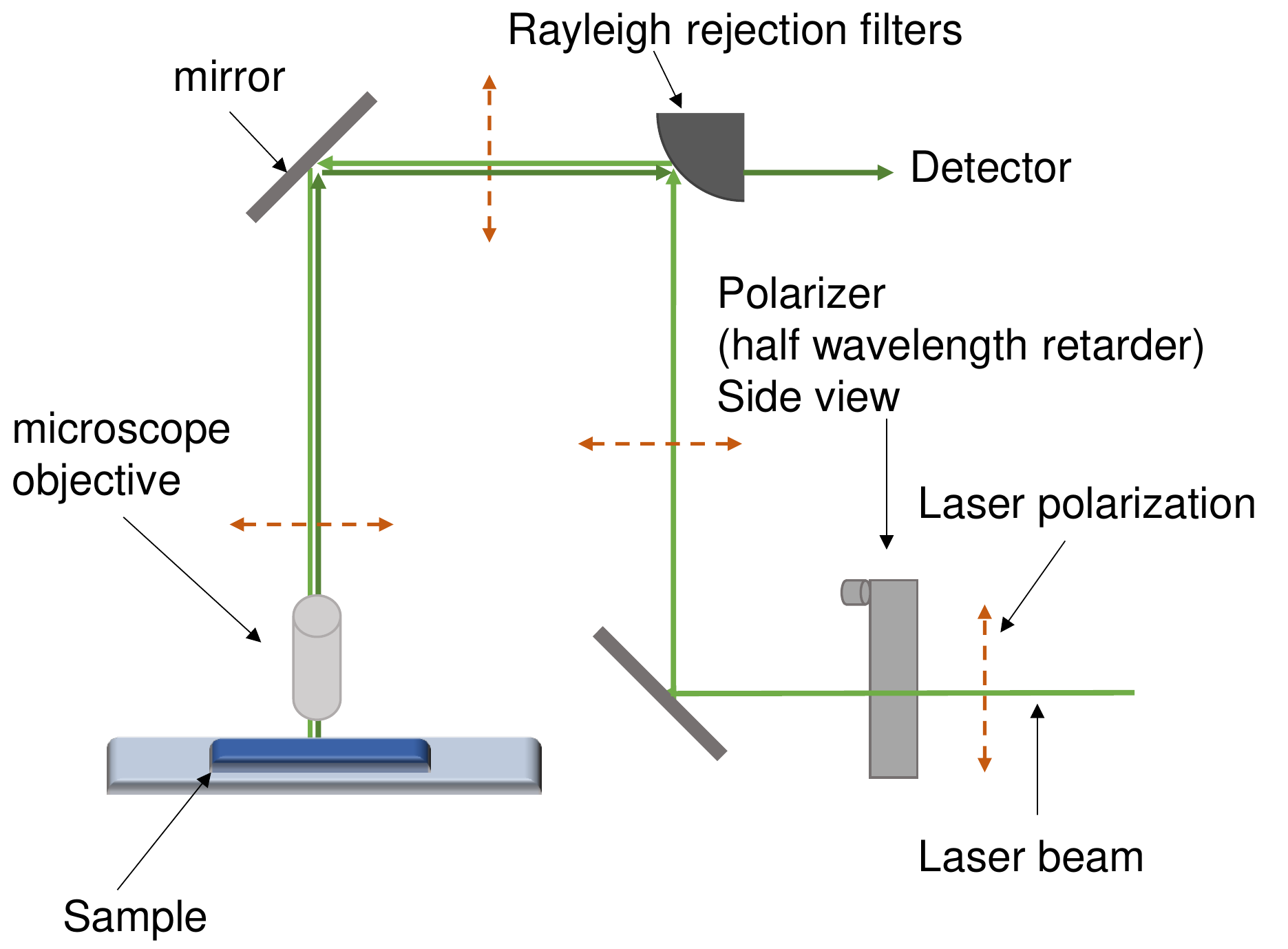}
   \caption{\label{Sfig:Raman-a}Diagram of the polarized Raman setup. The incoming laser beam goes through a half-wavelength retarder before hitting the sample. The sample is fixed and can be monitored between  measurements using the built-in optical microscope. The laser wavelength is 532~$\si{\nano\meter}$. The polarization of the laser source is linear and  is rotated in--plane through the half wavelength retarder. Detection occurs in a backscattering configuration.}
\end{figure}

The collected spectra are shown in figure~\ref{Sfig:Raman-b}. The three characteristic bP peaks, $A_g^1$, $B_{2g}$ and $A_g^2$ \cite{Sugai1985}, are clearly visible as well as their intensity modulation with the incoming laser beam polarization angle. The collected spectra underwent a linear background subtraction, obtained by fitting with a line the 10\%-15\% data in the beginning and in the end of each spectrum, where no peak is present.

\begin{figure}[tb]
   \includegraphics[scale=0.7]{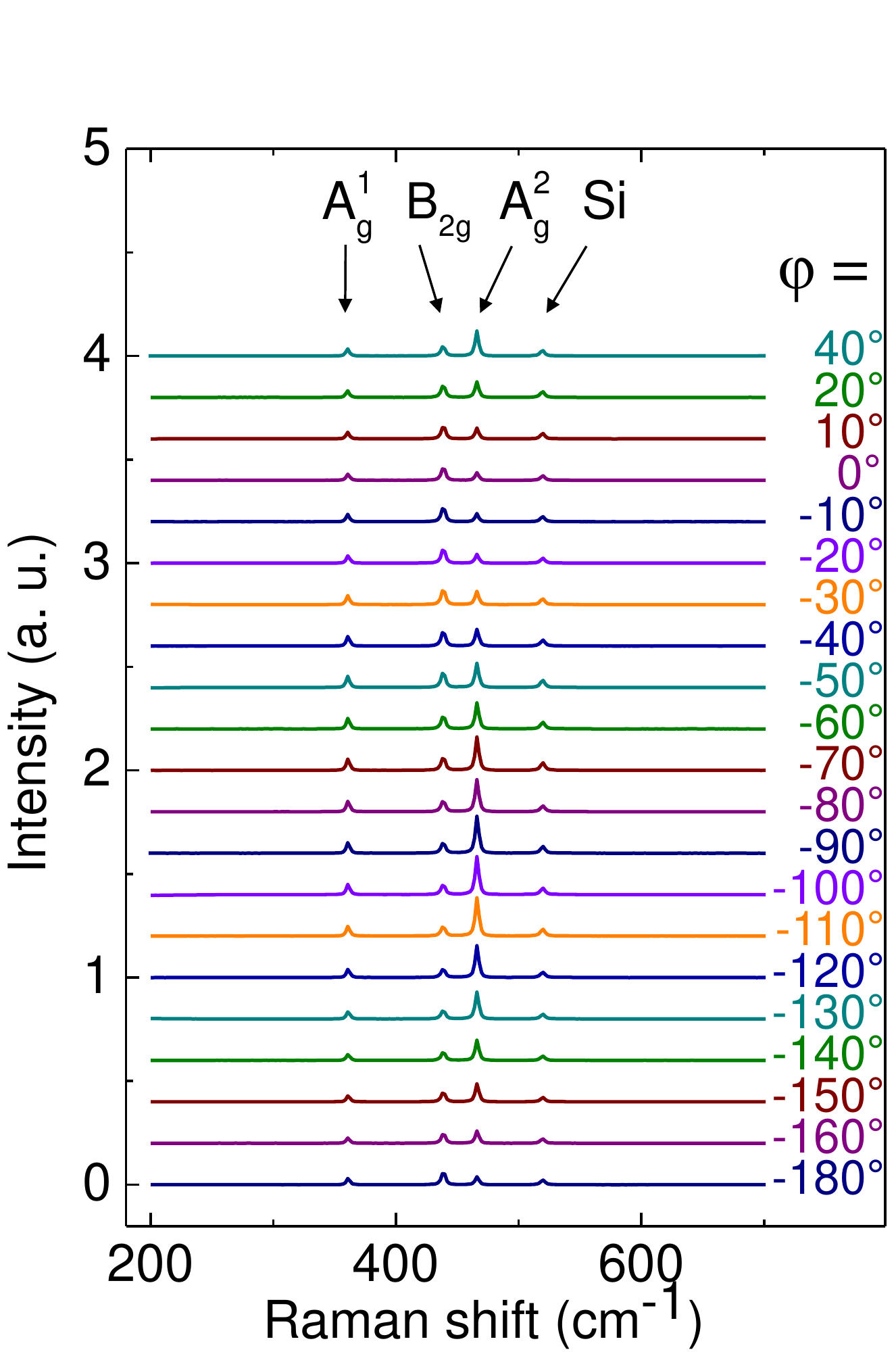}
   \caption{\label{Sfig:Raman-b}Raman spectra with baseline subtracted. The data traces are acquired at different polarization angles $\varphi$ as shown in the figure. The angle labeling convention is similar than in the main text. The data shown in the figure are shifted with a constant offset to allow a peak height evaluation. At all polarizations, the three Raman peaks of bP, namely $A_g^1$, $B_{2g}$ and $A_g^2$ are clearly visible, in addition to a peak related to the Si substrate (shown by arrows). The modulation of the bP Raman peaks intensity with polarization angle can be seen and it is particularly significant for the $A_g^2$ peak.}
\end{figure}

The ability of Raman spectroscopy to determine the crystallographic orientation of bP is well known \cite{Phaneuf-LHeureux2016a}. Using the $A^2_g$ peak is the most straightforward route to determine it, since it exhibits a maximum intensity along the armchair direction \cite{Sugai1985, Phaneuf-LHeureux2016a}. Previous work \cite{Kim2015a} used this technique to successfully determine the crystal orientation of bP for a flake of comparable thickness than ours and by using the same wavelength. In contrast, the intensity trend of the $B_{2g}$ peak is more complex. In cross-polarized Raman measurements it displays a four-lobe intensity versus angle  \cite{Phaneuf-LHeureux2016a, Kim2015a}, both for parallel and perpendicular polarization detected, with a shift in angle among the two. The $A_g^2$ peak is thus a better indicator of crystal orientation.
Since the peak wavenumber is well defined and the full with at half maximum (FWHM) is constant for each trace in the data set, the $A^2_g$ peak intensity can be simply identified as the maximum value of the peak in the acquired raw data.

This intensity versus $\varphi$ is shown in figure~1 of the main text and it is clearly sinusoidal. Fitting a sine function to these data, leaving both period and phase as free parameters (adjusted  $R^2>0.98$), a periodicity of $(87.2 \pm 1.1)^\circ$ and a center value of $(35.7 \pm 1.6)^\circ$ are obtained. This corresponds to the angular position of the intensity maximum, which is along armchair direction, at $-95.1^\circ$. Since armchair and zig-zag directions are orthogonal, this gives the zig-zag direction at $-5.1^\circ$.

Imposing a $90^\circ$ periodicity constraint on the sine fitting function for $A^2_g$, an equally good fit convergence is obtained (adjusted  $R^2>0.97$). The maximum of the $A^2_g$ peak, corresponding to the armchair axis, is at $-96.0^\circ$, and its minimum, which corresponds to the zig-zag direction, at $-6.0^\circ$. This is a shift of less than $1^\circ$ with respect to the unconstrained fit, well within our interval of confidence. It can therefore be stated that the zig-zag direction is at $(-5 \pm 3)^\circ$ with respect to the source-drain direction, see figure~1 of the main text. The $3^{\circ}$ error is an estimation that takes into account both the propagated error and the systematic error due to the possible orientation misalignment of the sample with respect to the Raman setup.

\section{Additional electronic transport characterization}

Magneto-transport measurements, shown in figure~3 and figure~4 of the main text, were performed at the National High Magnetic Field Laboratory in Tallahassee in a Helium-3 cryostat located in the  $45\,\si{\tesla}$ hybrid magnet. Electronic transport measurements were performed using standard ac lock-in techniques. A sourcing voltage of 1 V at 11.144 Hz with a resistance of 1 $\si{\mega\ohm}$ in series was used. As the sample resistance was non-negligible compared to that of the sourcing resistor, the exact current being sourced to the sample was weakly dependent on the sample resistance. This was corrected for in the analysis. A mechanical actuator was used to sweep the angle of the source-drain axis over more than $180^\circ$ with respect to the magnetic field in the sample plane, as shown in figure~2 of the main text.

\begin{figure}[tb]
   \includegraphics[scale=0.7]{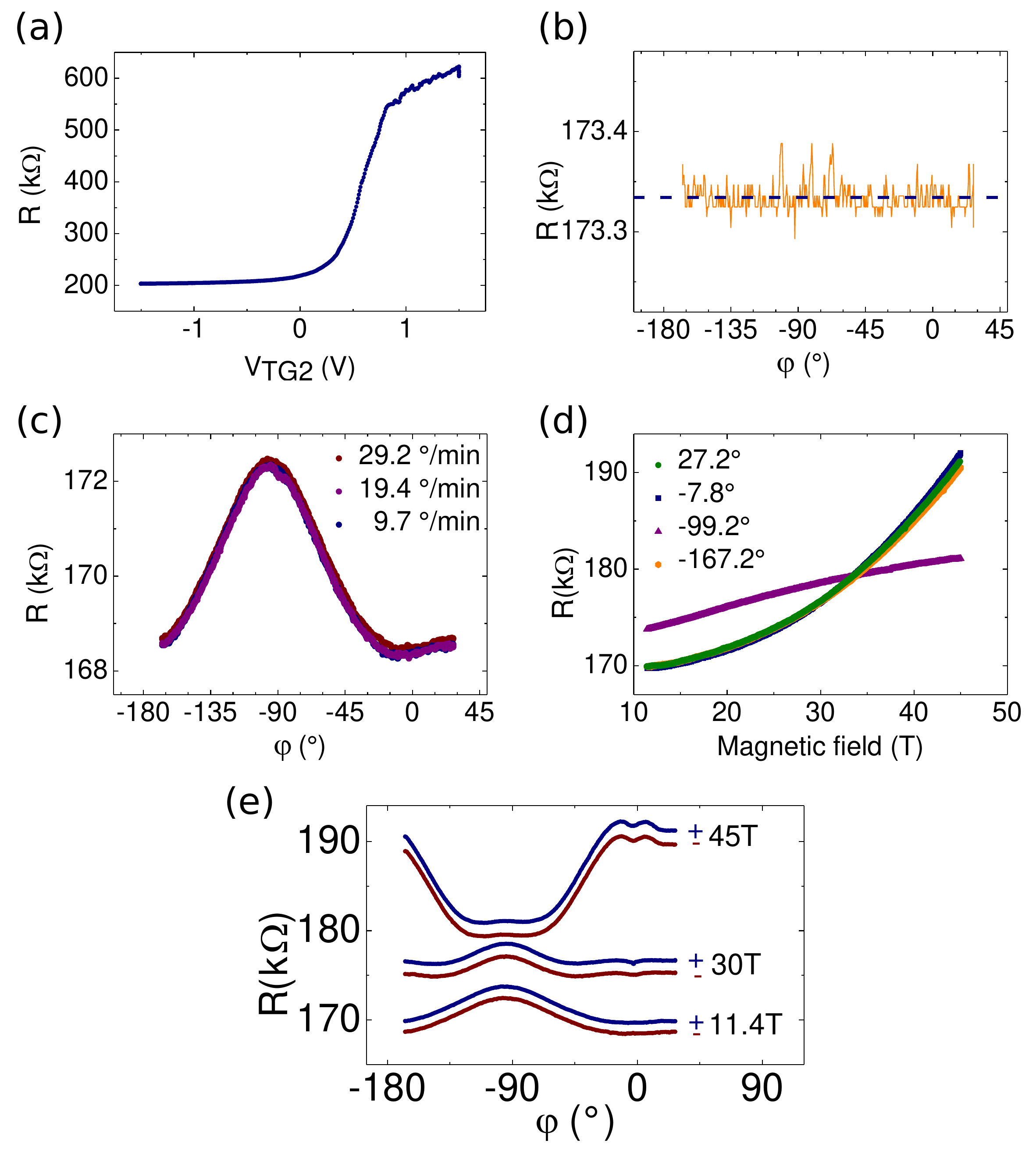}
   \caption{\label{Sfig:3}({\bf a}) Source-drain resistance $R$ as a function of $V_{TG2}$ at B=0~$\si{\tesla}$. This data were acquired during a different cooldown. ({\bf b}) Angular sweep at B=0~$\si{\tesla}$. The blue dashed line on the graph shows the average value that is used to normalize the data shown in Fig.~3 and 4 of the main text. ({\bf c}) Angular measurements for various sweep rates, namely 9.7$^\circ$/min, 19.4$^\circ$/min, and 29.2$^\circ$/min, acquired at $B=-11.4$~$\si{\tesla}$. ({\bf d}) Magnetoresistance measurements at fixed angles, namely -167.2$^\circ$, -99.2$^\circ$, -7.8$^\circ$ and 27$^\circ$.  ({\bf e}) Magnetoresistance for both polarity of the magnetic field. The two measurements (red - negative - and blue - positive) were acquired during two different cooldowns. 
In all panels except ({\bf a}), $V_\text{TG2}=-1\,\si{\volt}$. All data were acquired at $T\sim323$~$\si{\milli\kelvin}$.}
\end{figure}

A top gate sweep is shown in figure~\ref{Sfig:3}~(a). In the experimental conditions used for the angle--resolved magneto--transport measurements, at $V_{TG2}=-1$~$\si{\volt}$, the channel is conducting and well below the pinch-off point. The resistance value shown in figure~S4 (a) is approximately 15\% higher than the resistance values shown in figure~3 (b). This is most likely due to a slight charge density shift because the data were acquired during different cooldowns. A similar effect is also visible in  figure~\ref{Sfig:3}~(e) since the data at positive and negative magnetic fields were also taken before and after a temperature cycle.

Field-effect mobility and carrier concentration can be estimated from measurements with a fixed $back$ gate voltages of 0~$\si{\volt}$ and -25~$\si{\volt}$ at 11.4~$\si{\tesla}$, and at $1.5$~$\si{\kelvin}$. From linear interpolation of the conductance versus gate voltage \cite{Tsuno1999, Schroder2006}, the threshold voltage for the device of $V_T = 31$~$\si{\volt}$ is obtained. The threshold voltage is defined as the voltage at which the induced charge compensates the intrinsic doping, in our case related to acceptors \cite{Grosso}. The intrinsic carrier concentration can thus be extracted as $n=C_{ox} V_T/e$, where $e$ is the electron charge and $C_{ox}$ is the capacitance per unit of area of the $300$~$\si{\nano\meter}$ SiO$_2$ back gate oxide, equal to $11.5$~ $\si{\nano\farad\per\centi\meter\squared}$. A carrier density of $2.2 \times 10^{12}$~cm$^{-2}$ is obtained. Using the Drude model, the mobility can be estimated as $\mu = L/(W R n e)$, where $L$ and $W$ are the FET channel length and width, respectively, and R is the resistance of the device. The obtained value is  $83$~$\si{\centi\meter\squared\per\volt\per\second}$. To check for consistency, an angular sweep at zero magnetic field was performed and is shown in figure~\ref{Sfig:3} (b). Within the resolution of our experiment, no angular dependence is observed, thereby confirming the absence of possible spurious effects related to the rotation mechanism. The average resistance of the device is ($173334 \pm 13$)~$\si{\ohm}$. This value is used as the normalization for the MR data shown in figure~3 and figure~4 of the main text.

We also took data at different sweep rate, in order to rule out dynamical effects. As shown in figure~\ref{Sfig:3} (c), there is no significant difference between the various data sets, and therefore such dynamical effects can be excluded. As a further check for data consistency, experiments were performed at several \textit{fixed} angles, as a function of magnetic field, see figure~\ref{Sfig:3}~(d). These data taken at fixed angle are fully consistent with those shown in figure~3 of the main text.

To rule out any influence of self heating due to the mechanical rotation of the device, the measured valus of the thermometers located on the sample holder were carefully investigated. In spite of the absolute value of the thermometers being not fully reliable at high magnetic fields, the analysis of their relative values provides us confidence about the absence of any significant self-heating spurious effect. This is consistent with the fast thermal exchange occurring between the device, the rotator and the $^3He$ liquid bath. Moreover, the temperature stability as a function of time was evaluated by comparing measurements at similar magnetic field, but acquired at different times. Temperature measurements taken at different time in the day differed by less than 3\% ($9$~$\si{\milli\kelvin}$ at $\sim 323$~$\si{\milli\kelvin}$).

Given that a misalignment of $1$~$^{\circ}$ degree at $45$~$\si{\tesla}$ would result in a perpendicular field component of $0.78$~$\si{\tesla}$, the possible presence of this out-of-plane field was considered. Although the presence of such out-of-plane component is possible, it cannot itself give rise to the phenomena investigated here. In particular, a potential correction due to weak localization at high magnetic field should decrease the resistance, and therefore it cannot explain the positive longitudinal magnetoresistance observed at high field.

\end{widetext}

\end{document}